\documentclass[twocolumn,prl,aps,nobalancelastpage]{revtex4}

\usepackage{graphicx}
\usepackage{dcolumn}
\usepackage{bm}

\begin{document}

\title{The ground state of heavily-overdoped non-superconducting La$_{2-x}$Sr$_x$CuO$_4$}

\author{S. Nakamae$^1$, K. Behnia$^1$, N. Mangkorntong$^2$, M. Nohara$^2$, H. Takagi$^{2,3,4}$, S. J. C. Yates$^5$
and N. E. Hussey$^5$} \affiliation{$^1$LPQ (UPR5-CNRS), ESPCI, 10 Rue Vauquelin, F-75005 Paris, France.}
\affiliation{$^2$Department of Advanced Materials Science, Graduate School of Frontier Sciences, University of
Tokyo, Hongo 7-3-1, Bunkyo-ku, Tokyo 113-8656, Japan.} \affiliation{$^3$Institute of Physical and Chemical
Research (RIKEN), Wako 351-0198, Japan.} \affiliation{$^4$Correlated Electron Research Center, AIST, Tsukuba,
Japan.} \affiliation{$^5$H. H. Wills Physics Laboratory, University of Bristol, Tyndall Avenue, Bristol BS8 1TL,
U.K.}

\date{Submitted to Physical Review Letters, 9 December 2002}

\begin{abstract}
We report detailed thermodynamic and transport measurements for non-superconducting La$_{1.7}$Sr$_{0.3}$CuO$_4$.
Collectively, these data reveal that a highly-correlated Fermi-liquid ground state exists in
La$_{2-x}$Sr$_x$CuO$_4$ beyond the superconducting dome, and confirm that charge transport in the cuprates is
dominated at finite temperatures by intense electron-electron scattering.
\end{abstract}

\maketitle

The high-$T_c$ cuprates (HTC) have emerged as one of the most formidable challenges facing the theory of
correlated electrons in solids \cite{orenstein00}. In particular, transport properties of the normal state
present a number of uneasy paradoxes for the Fermi-liquid (FL) picture. Whilst it has been argued that
conventional electron-phonon ($e$-$ph$) scattering might still account for the $T$-linear resistivity in
optimally doped cuprates \cite{allen01}, it is generally assumed that proximity to a Mott insulator, and thus
the unusual strength of on-site electronic repulsion, is the fundamental reason behind the more unusual aspects
of HTC physics. This general assumption is backed up by a host of experimental observations which indicate that
the anomalous behavior becomes even more pronounced with decreasing the density of carriers, that is, by moving
towards the underdoped side. At sufficiently high carrier concentrations however, it has often been assumed that
HTC eventually evolve into a conventional FL as the electron correlations become weaker and the system becomes
more three-dimensional (3D).

Ironically, the persistence of robust superconductivity on the overdoped (OD) side of the phase diagram has been
a major obstacle in the exploration of the metallic non-superconducting ground state in HTC. Indeed, supporting
evidence for a FL ground state has only surfaced very recently with the experimental verification of the
Wiedemann-Franz (WF) law in OD Tl$_2$Ba$_2$CuO$_{6+\delta}$ (Tl2201) ($T_c$ $\sim$ 15K) \cite{proust02}. By
suppressing superconductivity in a large magnetic field, Proust {\it et al.} observed the precise WF ratio
$\kappa_{ab}$/$\sigma _{ab} T$ = $L_0$, where $\kappa _{ab}$ and $\sigma _{ab}$ are the in-plane thermal and
electrical conductivities and the Lorenz number $L_0$ = 2.44 $\times $ 10$^{-8}$ W$\Omega$/K$^{-2}$.
Surprisingly however, and at odds with a conventional FL picture, the WF relation was found to co-exist with a
large linear resistivity term extending down to 0K. This dichotomy raises the question whether the field-induced
'normal state' in OD HTC, i.e. beyond $H_{c2}$, is the same as the ground state that would exist in the absence
of a magnetic field, as it does in more conventional superconductors. Moreover, a clear understanding of the
experimental situation in OD cuprates has often been compounded by their tendency to undergo phase separation.

In this Letter, we present in- ($\rho_{ab}$) and out-of-plane ($\rho_c$) resistivity, in- ($\kappa_{ab}$) and
out-of-plane ($\kappa_c$) thermal conductivity, specific heat ($C$) and magnetic susceptibility ($\chi$)
measurements on single-phase non-superconducting La$ _{2-x}$Sr$_x$CuO$_4$ (LSCO) crystals ($x$ = 0.30) in which
these various concerns are removed. Collectively, the data provide a consistent picture of La$
_{1.7}$Sr$_{0.3}$CuO$_{4}$ as a highly-correlated FL. The WF law is verified to within our experimental
resolution, though in contrast to OD Tl2201, both $\rho_{ab}$ and $\rho_c$ exhibit strictly $T$$^2$ behavior
below 50K, {\it with no additional linear term}. Significantly, the Kadowaki-Woods ratio, linking the
coefficient $A$ of the (in-plane) $T$$^2$ resistivity and the square of the linear specific heat coefficient
$\gamma_0$, is found to be anomalously enhanced, even compared with other strongly correlated metals. This
latter observation reveals that intense $e$-$e$ interactions persist beyond the superconducting dome and sheds
important new light on the evolution of $\rho_{ab}$($T$) across the HTC phase diagram.

Seven bar-shaped samples (typical dimensions 3 mm x 0.6 mm x 0.6 mm) were prepared for either $ab$-plane (A1-4)
or $c$-axis (C1-3) measurements from a large La$_{1.7}$Sr$_{0.3}$CuO$_4$ single crystal grown in an infra-red
image furnace. The individual ingots were post-annealed together with the remaining boule under extremely high
partial pressures of oxygen (400 atm) for 2 weeks at 900$^\circ$C to minimize oxygen deficiencies and to ensure
good homogeneity within each crystal. Subsequent X-ray analysis revealed good crystallinity and no trace of
superconductivity could be detected (resistively) down to 95mK (see top inset to Fig. 1), confirming that these
crystals were indeed single-phase. $\chi$($T$) of the boule was measured with a commercial magnetometer, whilst
$C$($T$) was measured between 0.6K and 10K in a relaxation calorimeter. In both cases, addenda contributions
were measured independently and subtracted from the raw data. $\kappa$($T$) of A1, A2 and C1 were measured in a
dilution refrigerator using three RuO$_2$ chips employed as one heater and two thermometers. $\rho$($T$)
measurements were made on all seven samples using a conventional ac four-probe method. For all transport
measurements, current and voltage contacts were painted onto the crystals so as to short out any spurious
voltage drops from orthogonal components of the conductivity tensor. Uncertainty in the geometrical factor was
estimated to be $\sim$ 20$\%$. Finally, $c$-axis magnetoresistance $\Delta \rho _c$/$\rho _c$ data were taken on
C1 at the NHMFL in Florida.

\begin{figure}
\includegraphics[width=8.0cm,keepaspectratio=true]{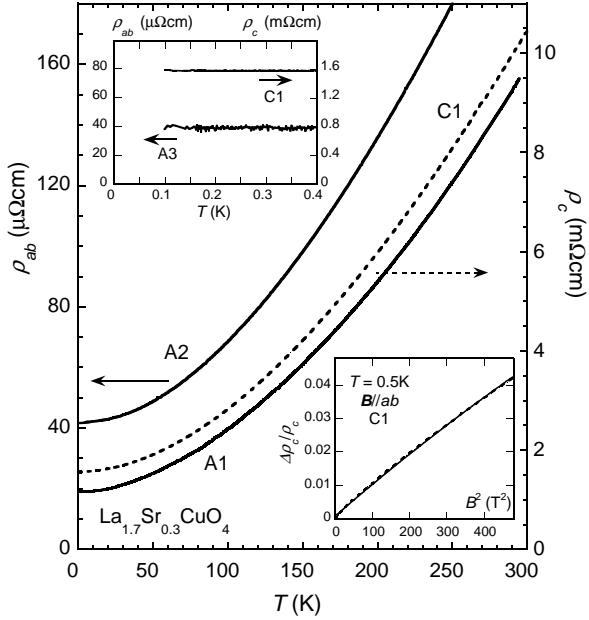}
\caption{Zero-field resistivity of A1, A2 ($\rho_{ab}$ - solid lines) and C1 ($\rho_c$ - dashed line). Upper
inset: Low-$T$ $\rho$($T$) of A3 and C1. Lower inset: $\Delta\rho_c$/$\rho_c$ versus $B$$^2$ for C1 at 0.5 K
($B\|ab$). The solid line is a fit to $\Delta\rho_c$/$\rho_c$ = 0.00010$B$$^2$ - 3.1 $\times$ 10$^{-8}$$B$$^4$.}
\label{1}
\end{figure}

Fig. 1 shows $\rho$($T$) of A1, A2 and C1 below 300K. (All crystals reported here showed similar behavior). Note
the similar metallic $T$-dependencies observed along both orthogonal directions, the different residual
resistivity ($\rho_0$) values for A1 and A2, and the strong upward curvature across the entire temperature
range. The resistivity ratio $\rho_c$(C1)/$\rho_{ab}$(A1) rises from $\sim $ 65 at 300K to $\sim$ 80 as $T
\rightarrow$ 0, presumably due to slight differences in their $\rho_0$ values. The limiting low-$T$
resistivities of C1 and A3 are reproduced in the top inset to Fig. 1. No current dependence was observed in
$\rho$($T$) down to 0.1A/cm$^2$, well below the typical critical current densities found in HTC. The lower inset
shows $\Delta \rho _c$/$\rho _c$ for C1 ($B\|ab$) plotted versus $B$$^2$ at $T$ = 0.5K. Inserting the fit to the
low-field data (see caption) into the Boltzmann transport equation for a quasi-2D FL (see, e.g.
\cite{hussey96}), we obtain an estimate of the in-plane electronic mean-free-path $\ell _{ab} \sim$ 145 $\AA $.
Finally, using $k_F$ = 0.55$\AA ^{-1}$ for La$_{1.7}$Sr$_{0.3}$CuO$_4$ \cite{ino02} and the ''isotropic-$\ell$''
approximation \cite{mackenzie96a}, we obtain $\rho _{ab0}$ = 21$\mu\Omega$cm, in good agreement with
$\rho_0$(A1).

\begin{figure}
\includegraphics[width=8.0cm,keepaspectratio=true]{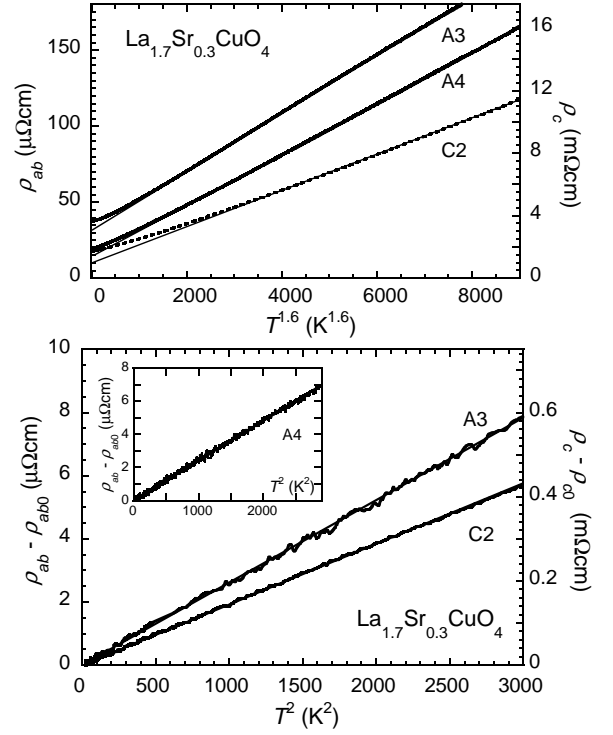}
\caption{a) $\rho$($T$) of A3, A4 (solid lines)
and C2 (dashed line) versus $T$$^{1.6}$. Thin solid lines are provided as guides. b) Low-$T$ $\rho$($T$) of A3
and C2 versus $T$$^2$. Inset: $\rho$ vs. $T$$^2$ for A4.}
\label{2}
\end{figure}

In the top panel of Fig. 2, $\rho$($T$) of A3, A4 and C2 are plotted versus $T$$^{1.6}$. In a previous study
\cite{takagi92}, such non-integer power-law resistivities were shown to extend from the lowest temperatures up
to 1000K. Whilst $\rho$($T$) of our crystals follows very closely a $T$$^{1.6}$ dependence at elevated
temperatures, the $T$-dependence clearly becomes stronger than $T$$^{1.6}$ as $T$ drops below around 100K.
Indeed, as shown in the bottom panel in Fig. 2, ($\rho$ - $\rho_0$)($T$) = $A$$T$$^2$ up to at least $T$ = 55K
{\it for both current directions} (($\rho$ - $\rho_0$)($T$) for A4 is shown in an inset for clarity).
Significantly, $A$ has the same magnitude ($\sim $ 2.5 $\pm$ 0.1 $n\Omega $cm/K$^2$) for all in-plane crystals,
even though $\rho_0$(A2,A3) $\sim$ 2 $\rho_0$(A1,A4). To our knowledge, this is the first time a pure $T$$^2$
resistivity has been resolved in a hole-doped cuprate (recall that in OD Tl2201, ($\rho - \rho_0$)($T$) =
$\alpha$$T$ + $AT$$^2$ \cite{proust02}) and implies that in La$_{1.7}$Sr$_{0.3}$CuO$_4$, charge transport can be
understood in terms of an anisotropic 3D FL \cite{2DFL}.

\begin{figure}
\includegraphics[width=8.0cm,keepaspectratio=true]{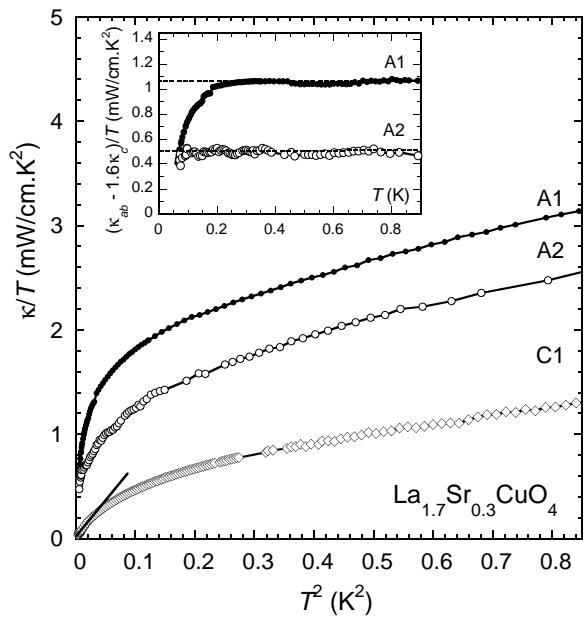}
\caption{Low-$T$ thermal conductivity of A1, A2 and C1 plotted as $\kappa/T$ vs. $T^{2}$. Inset: ($\kappa_{ab}$
- 1.6$\kappa_c$)/$T$ for A1 and A2. The dotted lines give $\kappa_{el}$/$T$ as $T \rightarrow$ 0.} \label{3}
\end{figure}

In Fig. 3, the low-$T$ thermal conductivities of A1, A2 and C1 are plotted as $\kappa$/$T$ versus $T$$^2$. Note
that A1 and A2 are better conductors of heat, as expected since $\sigma_c$ $\ll$ $\sigma _{ab}$, and that
$\kappa$(A1) $>$ $\kappa$(A2), reflecting their respective $\rho _0$ values. The sizeable phonon term
($\kappa_{ph}$) in HTC has often been a major obstacle to the determination of $\kappa _{el}$, the electronic
contribution to $\kappa$($T$) and the chief quantity of interest. Typically, $\kappa_{ph}$ can only be
unambiguously determined in the ballistic limit, below say 0.2K, where $\kappa_{ph}$ is simply proportional to
$\beta_3$$T$$^3$, the low-$T$ lattice heat capacity, as found here for C1 and indicated in the figure by a solid
line. If $\kappa_{ph}$ were isotropic, a more insightful approach might be to measure both $\kappa _{ab}$ and
$\kappa _{c}$ on identical crystals and make use of the large $\sigma _{ab}$/$\sigma _c$ anisotropy in HTC to
delineate $\kappa _{el}$ from $\kappa_{ab}$ across the entire temperature range.

A quick inspection of Fig. 3 however reveals that $\kappa_{ab}$/$T$ and $\kappa_c$/$T$ are not simply offset
from one another, implying either a $T$-dependent $\kappa_{el}$/$T$ at low-$T$ or anisotropy in $\kappa
_{ph}$($T$) for in- and out-of-plane heat flow, due, for example, to anisotropy in the phonon velocities or
$e$-$ph$ coupling. The fact though that $\kappa$/$T$(A1) and $\kappa$/$T$(A2) are shifted relative to one
another (except at the very lowest temperatures) suggests the latter, and simply by scaling $\kappa_c$($T$) by a
constant factor ($f$ = 1.6) and subtracting it from $\kappa_{ab}$($T$), we find, as shown in the inset to Fig.
3, that the resultant $\kappa_{el}$/$T$ is indeed constant for both crystals and equal to 1.07 and 0.50
mW/cm.K$^2$ for A1 and A2 respectively (the dotted lines). This gives $\rho_0$ = $L_0$$T$/$\kappa_{el}$ = 23
(A1) and 49 (A2) $\mu\Omega$cm, compared with the measured values of 19 and 42 $\mu\Omega$cm. Given our somewhat
simplistic way of extracting $\kappa_{ph}$($T$) from $\kappa_{ab}$($T$), it is encouraging to find that the WF
law is satisfied for both crystals to within our experimental (geometrical) error. It should be stressed here
that $\sigma_{ab0}$(LSCO) is almost an order of magnitude smaller than in OD Tl2201 \cite{proust02}, thus making
determination of $\kappa_{el}$/$T$ more difficult.

The surprising feature of Fig. 3 is of course the strong downturn in $\kappa$(A1) as $T \rightarrow$ 0. (A
slight downturn is also observed for A2 below 0.1K, though nowhere near as dramatic.) While slight deviations
from the WF law can be attributed to additional anisotropy in $\kappa _{ph}$ not yet accounted for, there is no
assumption on $\kappa _{ph}$($T$) which can erase the sudden vanishing of $\kappa_{el}$ below 0.2K for A1. Note
that this downturn is robust both to changes in the thermal contacts and the crystal dimensions. Intriguingly, a
vanishing of $\kappa_{el}$ has also been observed in the electron-doped cuprate Pr$_{2-x}$Ce$_x$CuO$_4$ (PCCO)
\cite{hill01}. By suppressing superconductivity with a high magnetic field, Hill {\it et al.} found that
$\kappa_{el}$/$T$, whilst violating the WF law for $T$ $>$ 0.3K (in the sense that it clearly exceeds
$L_0$/$\rho$($T$)), rapidly vanishes as $T$ falls below 0.3K. These two anomalous observations appear to have a
common origin. The sample-dependent aspect of the downturn, however, suggests that it may be extrinsic in origin
and unrelated to any exotic electronic behavior, particularly in view of the other, more conventional results
obtained here for La$_{1.7}$Sr$_{0.3}$CuO$_4$ above 0.2K. We reserve a detailed discussion on the origin of this
anomalous downturn to a more complete and systematic investigation \cite{paglione}.

The absence of superconductivity in La$_{1.7}$Sr$_{0.3}$CuO$_4$, coupled with the emergence of a $T$$^2$
resistivity at low $T$, gives us a unique opportunity to compare experimental manifestations of {\it e-e}
correlations in La$_{1.7}$Sr$_{0.3}$CuO$_4$ to other strongly correlated systems. Electronic correlations in a
FL are known to lead to an enhancement in the quasi-particle effective mass. This effect can be detected through
the resistivity ($AT$$^2$), specific heat ($\gamma_0$$T$) and magnetic susceptibility ($\chi _{p}$). Empirical
relationships that correlate these physical parameters have been found in a wide range of strongly-correlated
metals; namely, the Kadowaki-Woods (KW) ratio ($A/\gamma_0^2$ = $a_0$ = 10$^{-5}\mu \Omega$cm/(mJ/Kmol)$^2$
\cite{kadowaki86,miyake89}) and the Wilson ratio ($R_w$ = $(\frac{\pi k_B^2}{3\mu _B^2}$) $\chi_p$/$\gamma_0$,
with $R_w \sim$ 1 for a free electron gas and $\sim $ 2 for strongly-correlated fermions \cite{maeno97}).

\begin{figure}
\includegraphics[width=8.0cm,keepaspectratio=true]{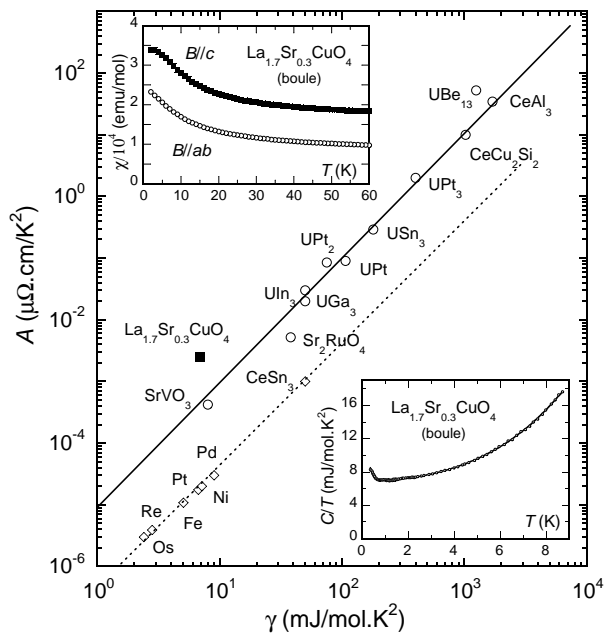}
\caption{Kadowaki-Woods plot of $A$, the coefficient of the $T$$^2$ resistivity, versus $\gamma_0$, the
electronic specific heat coefficient, for a variety of strongly correlated metals (adapted from
\cite{miyake89,maeno97,inoue98}). Upper inset: Magnetic susceptibility $\chi$($T$) of large boule at 5T with
$B\|c$ (solid squares) and $B\|ab$ (open circles). Lower inset: Low-$T$ specific heat of boule plotted as
$C$/$T$ vs. $T$$^2$. The solid line is a fit to the expression $C$ = $\alpha T$$^{-2}$ + $\gamma_0$$T$ +
$\beta_3 T$$^3$ + $\beta_5 T$$^5$. See text for parameter values.} \label{4}
\end{figure}

The lower inset to Fig. 4 shows $C$($T$) of the large crystal boule plotted as $C$/$T$ versus $T$$^2$. $C$($T$)
can be fitted over the whole temperature range (0.6K $<$ $T$ $<$ 10K) to the expression $C$ = $\alpha T$$^{-2}$
+ $\gamma_0 T$ + $\beta_3 T$$^3$ + $\beta_5 T$$^5$ with $\alpha$ = 76 $\mu$J.K/mol., $\gamma_0$ = 6.9
mJ/mol.K$^2$, $\beta_3$ = 0.1 mJ/mol.K$^4$ and $\beta_5$ = 0.7 $\mu$J/mol.K$^6$. The $\alpha T$$^{-2}$ term
represents the high temperature tail of a Schottky anomaly that invariably develops in HTC samples at low $T$
due to either a small concentration of isolated paramagnetic impurities, or to nuclear hyperfine or quadrupole
splitting. The phononic $\beta_3 T^3$ term is comparatively small and gives rise to an elevated $\Theta_D$ =
550K, though $\beta_3$ may be made artificially low by a $T$-dependent electronic term \cite{cooper96}.

The magnitude of $\gamma_0$, though, is very robust in our measurements, is comparable to previous reports at
this doping level \cite{cooper96} and gives a corresponding KW ratio, $A/\gamma_0^2 \sim$ 5$a_0$. This puts
non-superconducting LSCO well off the so-called 'universal' line for strongly correlated metals indicated in
Fig. 4. A similarly enhanced KW ratio, i.e. $A/\gamma_0^2 \geq$ 5$a_0$, can also be inferred for OD Tl2201 ($A$
= 5.4 n$\Omega$cmK$^{-2}$ \cite{proust02}, $\gamma_0$ $\sim$ 10 mJ/mol.K$^2$ \cite{wade95}), and indirectly for
OD PCCO, ($A$ $\sim$ 4 n$\Omega$cmK$^{-2}$ \cite{fournier98}, $\gamma_0$ $\sim$ 6.7 mJ/mol.K$^2$
\cite{balci02}). This observation reveals a new aspect of the physics of the cuprates and is the central result
of this Letter. Whilst various explanations for the magnitude of the KW ratio in heavy fermion systems have been
proposed, including a strong frequency dependence of the quasi-particle self-energy \cite{miyake89} and
proximity to an antiferromagnetic instability \cite{takimoto96}, none have so far predicted deviations from
$a_0$ as large as those found in the HTC. It is noted, however, that the effect of a strong momentum dependence
in the scattering amplitude, thought to be a key feature of HTC \cite{hussey96}, has yet to be taken into
account \cite{okabe98} and efforts to include such an effect could prove highly illuminating.

As one moves across the HTC phase diagram towards the Mott insulator at half-filling, one expects $e$-$e$
scattering to become even more intense. Thus, the gradual evolution from quadratic to linear resistivity in HTC
must somehow reflect the growing strength of $e$-$e$ interactions as one approaches the Mott insulator from the
metallic side. As an indication of how rapidly the $e$-$e$ scattering intensity might grow, we note here that
$\rho_{ab}$ of La$_{1.85}$Sr$_{0.15}$CuO$_4$ is $\approx$ 2 - 3 times larger than that of
La$_{1.7}$Sr$_{0.3}$CuO$_4$ at 300K, even though the Luttinger sum rule dictates that their carrier densities
differ by only a few percent. Moreover, given the relatively minor changes observed in the phonon spectra as a
function of doping, one does not expect the strength of the $e$-$ph$ interaction to change forcibly across the
phase diagram. Thus, since $e$-$ph$ scattering appears to give a negligible contribution to $\rho_{ab}$($T$) in
La$_{1.7}$Sr$_{0.3}$CuO$_4$, it seems highly unlikely now that $e$-$ph$ scattering can in any way account for
the linear resistivity that appears (in a narrow composition range) around optimal doping. In the light of these
results, we conclude that chronic (Umklapp) $e$-$e$ scattering processes must dominate the normal state
transport behavior of the cuprates across the \emph{entire} accessible doping range. Such a scenario is
consistent with the large increase in the quasi-particle lifetime below $T_c$ observed in thermal and electrical
conductivity measurements \cite{yu92,bonn93}.

Finally, let us comment briefly on the Wilson ratio $R_W$. As shown in the upper inset to Fig. 4, $\chi$($T$)
displays significant anisotropy with respect to the field orientation (believed to arise from anisotropic
$g$-values \cite{terasaki92}) and a strong enhancement at low $T$. Assuming a constant Pauli susceptibility
$\chi_p$ \cite{vv} and an isotropic Curie-Weiss term (with $\Theta_C$ $\sim$ 7.5K), we obtain an average
$\chi_p$ $\sim$ 1.75$\times$ 10$^{-4}$ emu/mol and hence, $R_W \sim$ 2.0, consistent with values for other
strongly correlated metals \cite{maeno97}. If however, all the enhancement in $\chi$($T$) is assumed to be
intrinsic, i.e. due to an enhanced $\chi_p$($T$), we obtain $R_W$ $\sim$ 3.5. We note here that whilst a large
enhancement in $\chi_p$($T$) at low-$T$ would imply a similar enhancement in $\gamma$($T$), such behavior would
be difficult to discern from our $C$($T$) data due to the dominant phonon contribution.

In summary, detailed low-$T$ transport and thermodynamic measurements in La$_{1.7}$Sr$_{0.3}$CuO$_4$ have
revealed for the first time that LSCO develops a highly correlated FL ground state beyond the superconducting
dome. The observed enhancement in the KW ratio suggests that $e$ scattering in non-superconducting LSCO is much
more intense than previously imagined and raises serious question marks over previous claims that the ubiquitous
linear resistivity observed in optimally-doped cuprates is a signature of (strong) $e$-$ph$ coupling.

We acknowledge stimulating discussions with Y. Ando, J. W. Loram, L. Taillefer and J. A. Wilson and experimental
assistance from L. Balicas. This work was supported by the Alliance: Franco-British Partnership Program.

\end{document}